\documentclass[prb,aps,onecolumn,preprint]{revtex4-1}

\usepackage{amssymb,amsmath}
\usepackage{graphicx}%Package to add figures with references
\usepackage{subcaption}%Package to combine two figures in one subfigure

\begin{document}

\title{Verification of an analytic fit for the vortex core profile in superfluid Fermi gases}

\author{Nick Verhelst}
\author{Sergei Klimin}
\author{Jacques Tempere}
\address{TQC, Universiteit Antwerpen, Universiteitsplein 1, B-2610 Antwerpen, Belgium}

\begin{abstract}
A characteristic property of superfluidity and -conductivity is the presence of quantized vortices in rotating systems. To study the BEC-BCS crossover the two most common methods are the Bogoliubov-De Gennes theory and the usage of an effective field theory. In order to simplify the calculations for one vortex, it is often assumed that the hyperbolic tangent yields a good approximation for the vortex structure. The combination of a variational vortex structure, together with cylindrical symmetry yields analytic (or numerically simple) expressions.

The focus of this article is to investigate to what extent this analytic fit truly reflects the vortex structure throughout the BEC-BCS crossover at finite temperatures. The vortex structure will be determined using the effective field theory presented in [Eur. Phys. Journal B \textbf{88}, 122 (2015)] and compared to the variational analytic solution. By doing this it is possible to see where these two structures agree, and where they differ. This comparison results in a range of applicability where the hyperbolic tangent will be a good fit for the vortex structure.
\end{abstract}

\maketitle

%\begin{keyword}
%Vortex structure \sep Imbalanced Fermi gases \sep Effective field theory

%\PACS 67.25.dk \sep 67.30.he \sep 67.85.-d \sep 03.75.Ss \sep 03.75.Lm


%\end{keyword}



%%%%%%%%%%
%Section 1: Introducing the article
%%%%%%%%%%
\section{Introduction: Vortices in the BEC-BCS crossover}
\label{sec:Introduction}
Quantized vortices are a hallmark for superfluidity and superconductivity, and have been a subject of interest since a long time \cite{BDZ2008}. Stable vortices and vortex arrays have been successfully realized in rotating condensates of bosons \cite{MAH1999,MCW2000,RAV2001,AS2001} and fermions \cite{ZAS2005}. Superfluid Fermi gases are of particular interest because the tunability of the interatomic interaction strength allows to investigate the crossover between a Bose-Einstein condensate (BEC) of strongly bound molecules and a Bardeen-Cooper-Schrieffer (BCS) state of Cooper pairs. The experimental achievements stimulated theorists to explore the physics of vortices and vortex matter in rotating, trapped quantum gases.

Different theoretical models can be applied to describe vortices. For Bose gases, the most common method is to employ the Gross-Pitaevskii (GP) equation \cite{TKU2002,FET2009}. Superfluid Fermi gases can however be studied by a variety of methods, the most common are: the Ginzburg-Landau (GL) formalism \cite{BV2001}, the Bogoliubov-De Gennes (BdG) theory \cite{MK2005, SRH2006, CCL2006, SPS2013,WS2011,WAR2012,WM2012}, superfluid density functional theory \cite{BUL2013}, the density matrix renormalization group \cite{YOM2008} and the coarse-grained BdG approximation \cite{SPS2015}.

To describe vortices in condensates throughout the BEC-BCS crossover, it appears that the BdG theory is the preferred method \cite{WS2011,WAR2012,WM2012}. The problem with the BdG theory is however that the method is computationally fairly cumbersome. Consequently the use of the BdG theory is mainly limited to the consideration of zero-temperature properties of single-vortex states  \cite{MK2005,SRH2006,CCL2006}. Because of this big computational cost of the BdG theory, there is a recent interest in the development of effective field theories \cite{SS2014,KTD2014a,KTLD2015,KTD2014,LAKT2015,NS2006,SCH2011}. These effective field theories allow for a description of non-uniform excitations (e.g. vortices and solitons) in finite-temperature Fermi gases throughout the BEC-BCS crossover. They require much less computational cost with respect to the BdG calculations and allow for the variational methods and sometimes for exact analytic solutions \cite{KTD2014}. 

Consequently, the effective KTD theory \cite{KTD2014a,KTLD2015,KTD2014} is used in the present work. The effective KTD theory corresponds nicely with the numerical BdG results, except in the deep BCS regime for temperatures far below $T_C$ \cite{LAKT2015}. The range of the considered scattering lengths $a_s$ will be limited to $(k_Fa_s)^{-1}\in[-1,2]$, where the effective KTD theory has a good correspondence.

In this paper, we use the KTD energy functional to study the order parameter 
in the neighbourhood of the vortex core, throughout the BEC-BCS crossover at finite temperatures. A common (variational) assumption is that the order parameter $\Psi$ heals according to $\Psi(r)=\Psi_{\infty} \tanh[r/(\sqrt{2}\xi)]$ where $r$ is the 
distance to the vortex core, $\xi$ is the characteristic ''healing'' 
length mentioned above, and $\Psi_{\infty}$ is the ''bulk'' order parameter far
away from the vortex. Here, we investigate how good the assumption of a
tanh-dependence is for a Fermi superfluid, in which regime the largest
deviations from it are to be expected, and how the resulting
estimate for the healing length is affected.

%%%%%%%%%%
%Section 2: Explaining the KDT theory
%%%%%%%%%%
\section{The effective field theory for vortices}
\label{sec:Theory}
The considered KTD effective field theory \cite{KTD2014,KTLD2015,LAKT2015} is derived using the path-integral formalism. The starting point is the Lagrangian for an $s$-wave scattering potential, which is a common low-temperature potential for atomic gases. The non-linear interaction term is then eliminated by using a Hubbard-Stratonovich transformation \cite{STRA1958,HUBB1959}, leading to a bosonic (pair) field $\Psi(\textbf{r},t)$ with an effective potential. Finally a gradient expansion (up to second order) is made around the coordinate-dependent saddle-point value of the bosonic field. This yields Matsubara summations which can be done analytically, resulting in the effective field theory. To allow for spin imbalance, chemical potentials $\mu_\sigma$ are introduced which can differ for the ''spin-up'' and ''spin-down'' species. Using these, the (average) chemical potential $\mu=(\mu_\uparrow+\mu_\downarrow)/2$ and (spin-)imbalance $\zeta=(\mu_\uparrow-\mu_\downarrow)/2$ are defined. In what follows, we use units $\hbar=2m=k_B=k_F=1$, with $2m$ the mass of a fermion pair and $k_F=[3\pi^2(N_\uparrow+N_\downarrow)/V]^{1/3}$ the Fermi wave vector. Note that, due to our choice of units, the total (pair) density is $n=1/(3\pi^2)$.

To introduce the vortex structure in a bulk medium, the bosonic pair field is written in polar coordinates $(r,\phi,z)$ as:
\begin{equation}
\Psi(\textbf{r})=\Psi_\infty f(r)e^{i\phi},
\label{eq:BosonicPairField}
\end{equation}
where $\Psi_\infty$ is the bulk-value of the pair field and $f(r)$ describes the vortex core profile. Rather than choosing a tanh-dependence for $f$, we find a numerical solution within the KTD effective field theory, and compare the two solutions. The profile function $f$ is subject to the boundary conditions $f(0)=0$ and $f(\infty)=1$. Adding the effects of rotation and substituting the vortex structure \eqref{eq:BosonicPairField} results in an effective energy given by \cite{KTD2014a}:
\begin{equation}
F_{eff}=\int d^3\textbf{r}\left[\Omega_s(|\Psi|^2)-\Omega_s(|\Psi_\infty|^2)+\frac{\rho_{sf}}{2r^2}(f(r))^2\right.\left.+\frac{\rho_{qp}}{2}(\partial_r f)^2\right].
\label{eq:EffectiveEnergy}
\end{equation}
In this expression, $\Omega_s$ is the thermodynamic grand potential per unit volume at inverse temperature $\beta=1/(k_BT)$, given by:
\begin{equation}
\Omega_s(|\Psi|^2)=-\int\frac{d\textbf{k}}{(2\pi)^3}\left(\frac{1}{\beta}\ln[2\cosh(\beta E_k)+2\cosh(\beta\zeta)]\right.\left.-\xi_k-\frac{|\Psi|^2}{2k^2}\right)-\frac{|\Psi|^2}{8\pi a_s},
\label{eq:GrandPotential}
\end{equation}
With $a_s$ the fermion-fermion scattering length, $\xi_k=k^2-\mu$ the free particle energy and $E_k=\sqrt{(k^2-\mu)^2+|\Psi|^2}$ the Bogoliubov excitation energy. The bulk superfluid density $\rho_{sf}$ and quantum pressure $\rho_{qp}$ are given by:
\begin{equation}
\begin{aligned}
\rho_{sf}&=2C(|\Psi_\infty|^2)|\Psi_\infty|^2,\\
\rho_{qp}&=2|\Psi_\infty|^2\left[C(|\Psi_\infty|^2)-4f(r)^2 E(|\Psi_\infty|^2)\right].
\end{aligned}
\label{eq:SupFluidQuantPress}
\end{equation}
The two coefficients $C$ and $E$ of the effective field theory are given by the integrals
\begin{equation}
\begin{aligned}
C(|\Psi_\infty|^2)&=\frac{2}{3}\int\frac{d\textbf{k}}{(2\pi)^3}k^2f_2(\beta,E_k,\zeta),\\
E(|\Psi_\infty|^2)&=\frac{4}{3}\int\frac{d\textbf{k}}{(2\pi)^3}k^2\xi_k^2f_4(\beta,E_k,\zeta),
\end{aligned}
\label{eq:Coefficients}
\end{equation}
written in terms of the functions $f_n(\beta,\epsilon,\zeta)$, which are defined recursively as:
\begin{equation}
\begin{aligned}
f_n(\beta,\epsilon,\zeta)&=\frac{1}{2\epsilon}\frac{\sinh(\beta\epsilon)}{\cosh(\beta\epsilon)+\cosh(\beta\zeta)}\\
 f_{n+1}(\beta,\epsilon,\zeta)&=-\frac{1}{2n\epsilon}\frac{\partial}{\partial\epsilon}f_n(\beta,\epsilon,\zeta).
\end{aligned}
\label{eq:CoefficientFunctions}
\end{equation}

In order to use the free energy functional \eqref{eq:EffectiveEnergy}, one first determines the bulk properties (without a vortex). This is done by simultaneously solving the gap and number equations which determine $|\Psi_\infty|$ and $\mu$ as a function of $1/(k_Fa_s)$ and $\beta$. Once these quantities are found, we minimize \eqref{eq:EffectiveEnergy} with respect to $f(r)$ in order to find the vortex core profile. Since \eqref{eq:EffectiveEnergy} is the result of a gradient expansion up to second order in the gradients of $\Psi$, it suffices to use $\Psi_\infty$ instead of $\Psi$ in the arguments of $C$ and $E$, as the difference is of higher order in gradients of $\Psi$.

%%%%%%%%%%
%Section 3: Explaining our variational and numerical methods
%%%%%%%%%%
\section{Methods}
\label{sec:Methods}
\subsection{Tanh-profile}
\begin{figure}[!t]
\centering
\includegraphics[width=0.7\linewidth]{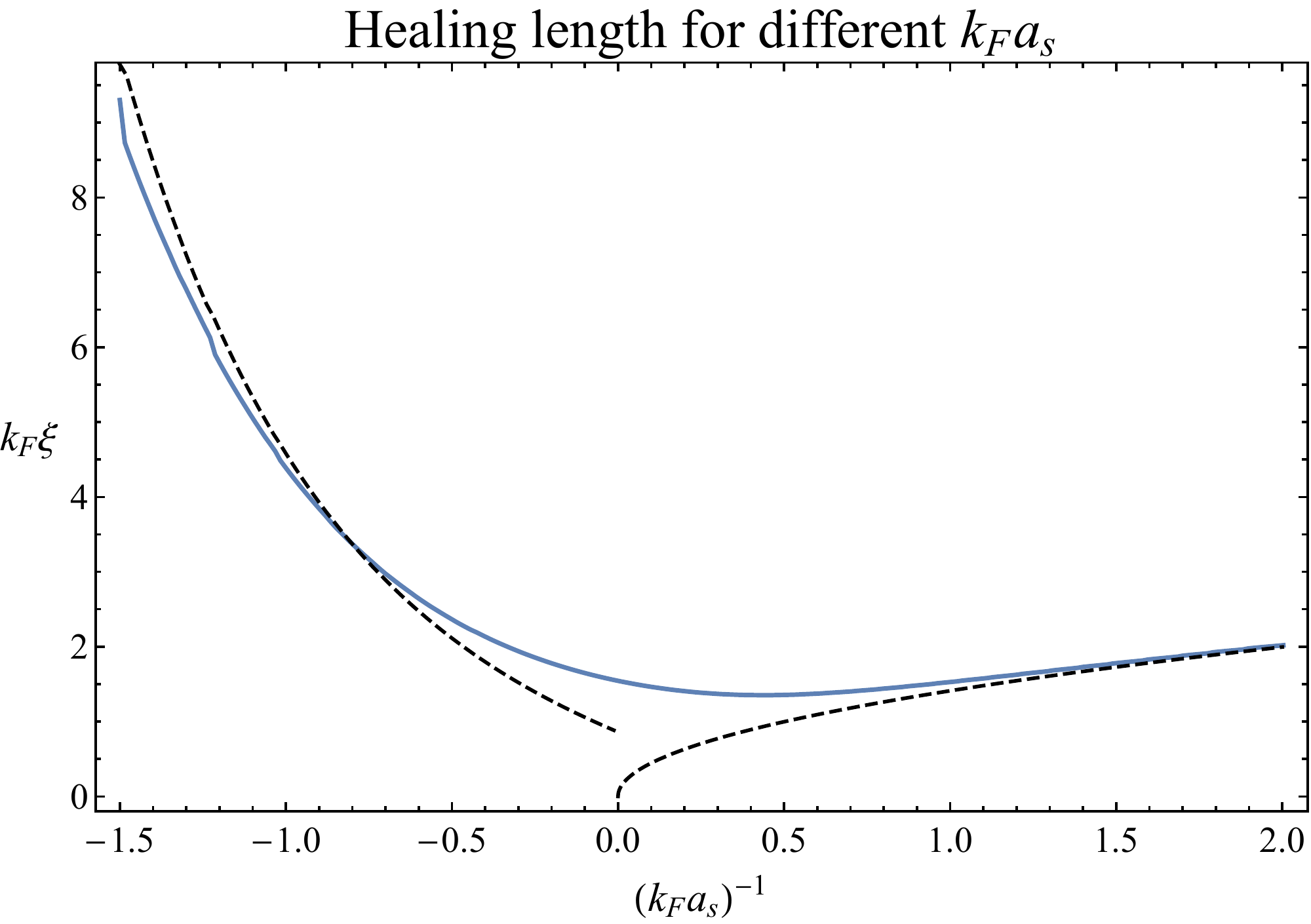}
\caption{The healing length $\xi$ throughout the BEC-BCS crossover for $\beta=100$ and $\zeta=0$. The dashed lines are the analytical results in the BEC and BCS limits.}
\label{fig:HealingLength}
\end{figure}

A common variational choice for the vortex profile is given by $f(r)=\tanh(r/(\sqrt{2}\xi))$, where $\xi$ is a variational parameter representing the healing length. The advantage of this  variational procedure is that the minimization can be performed analytically, yielding
\begin{equation}
\xi=\frac{1}{2}\sqrt{\frac{\rho_{sf}}{A}},
\label{eq:HealingLengthSergei}
\end{equation}
with
\begin{equation}
A=\int\limits_0^\infty r \left[\Omega_s\left(|\Psi_\infty|^2\tanh^2\left(\frac{r}{\sqrt{2}}\right)\right)-\Omega_s\left(|\Psi_\infty|^2\right)\right]dr.
\label{eq:CoefficientA}
\end{equation}
Using \eqref{eq:HealingLengthSergei} the healing length can be calculated throughout the BEC-BCS crossover, for $\beta=100$ and $\zeta=0$ we obtain figure \ref{fig:HealingLength}. In the BEC limit, $1/(k_Fa_s)\rightarrow\infty$, as well as in the BCS limit, $1/(k_Fa_s)\rightarrow-\infty$, we get a good agreement with the known analytic results \cite{Pita1961,MPS1998} for the coherence length, indicated as dashed curves in figure \ref{fig:HealingLength}.

\subsection{General profile}
To calculate the vortex profile $f(r)$ without resorting to a variational model such as the tanh-dependence, we perform a functional minimization of \eqref{eq:EffectiveEnergy} for a general function $f(r)$. The first step is to introduce a grid for the numerical representation of $f(r)$. Since \eqref{eq:EffectiveEnergy} only depends on the distance $r$ to the vortex line, the integrals over the polar coordinates $\phi$ and $z$ can be done analytically. As a large-r cutoff $R_c$ for the grid, we take twenty times the healing length\footnote{The results already become stable around ten to fifteen times the healing length.} of the hyperbolic tangent solution \eqref{eq:HealingLengthSergei}. Writing the integrand of the free energy \eqref{eq:EffectiveEnergy} as $\mathcal{F}(r)$, the discretization yields
\begin{equation}
\frac{F_{eff}}{2\pi H}=\int\limits_0^{R_c} r\mathcal{F}(r)dr\approx\sum_{n=2}^{N+1}\left[r_n\mathcal{F}(r_n)\right](r_n-r_{n-1}),
\label{eq:LatticeFreeEnergy}
\end{equation}
where the grid of $r$-points is given by $\lbrace r_1=0,r_2,\cdots,r_{N+1}=20\xi\rbrace$. Since the vortex profile varies more strongly near the origin, the sampling is chosen in such a way that 90\% of the points will lie in the interval $r\in[0,10\xi]$. For the derivatives, a backwards differentiation scheme is chosen, using this, the free energy density becomes:
\begin{equation}
\mathcal{F}(r_n)=\Omega_s(f^2_n|\Psi_\infty|^2)-\Omega_s(|\Psi_\infty|^2)+\frac{\rho_{sf}}{2r_n^2}f_n^2+\frac{\rho_{qp}}{2}\left(\frac{f_n-f_{n-1}}{r_n-r_{n-1}}\right)^2,
\label{eq:FreeEnergyDensity}
\end{equation}
with $f_n=f(r_n)$.

To find the true vortex structure, the free energy \eqref{eq:LatticeFreeEnergy} should now be minimized with respect to the set of variational parameters $\lbrace f_2,...,f_N\rbrace$, where the boundary conditions $f_1=0$ and $f_{N+1}=1$ are imposed. This minimization was done by a Monte-Carlo type algorithm. In order to have a fast convergence the initial guess for the values of $f_n$ are given by\footnote{The calculations were also done with a random scatter between 0 and 1, this yields the same final results for the algorithm.}
\[
f^{(0)}_n=\tanh\left(\frac{r_n}{\xi\sqrt{2}}\right),
\]
where $\xi$ is calculated with \eqref{eq:HealingLengthSergei}, $f^{(0)}_1=0$ and $f^{(0)}_{N+1}=1$. The upper index in $f_n^{(i)}$ indicates that $f_n^{(i)}$ is the $i$-th iterative Monte-Carlo approximation to the true profile function. This iteration goes as follows: the numerical algorithm runs sequentially through the list of $f_n^{(i)}$ values, from $n=2$ up to and including $n=N$. For each value of $f_n^{(i)}$ two new (random) values are generated as $f_n^{i,\pm}=(1+\delta_0\mathrm{RAND}[0,1])f_n^{(i)}$, where $\mathrm{RAND}[0,1]$ is a random number between 0 and 1, and $\delta_0=10^{-m}$ with $m\in\mathbb{N}$. The energy of the old structure is compared with the energy of the new structure where $f_n^{(i)}$ is replaced by $f_n^{i,\pm}$, the value with the lowest energy will be chosen for the new function value $f_n^{(i+1)}$.

\begin{figure}[!htb]
\centering
\includegraphics[width=0.45\linewidth]{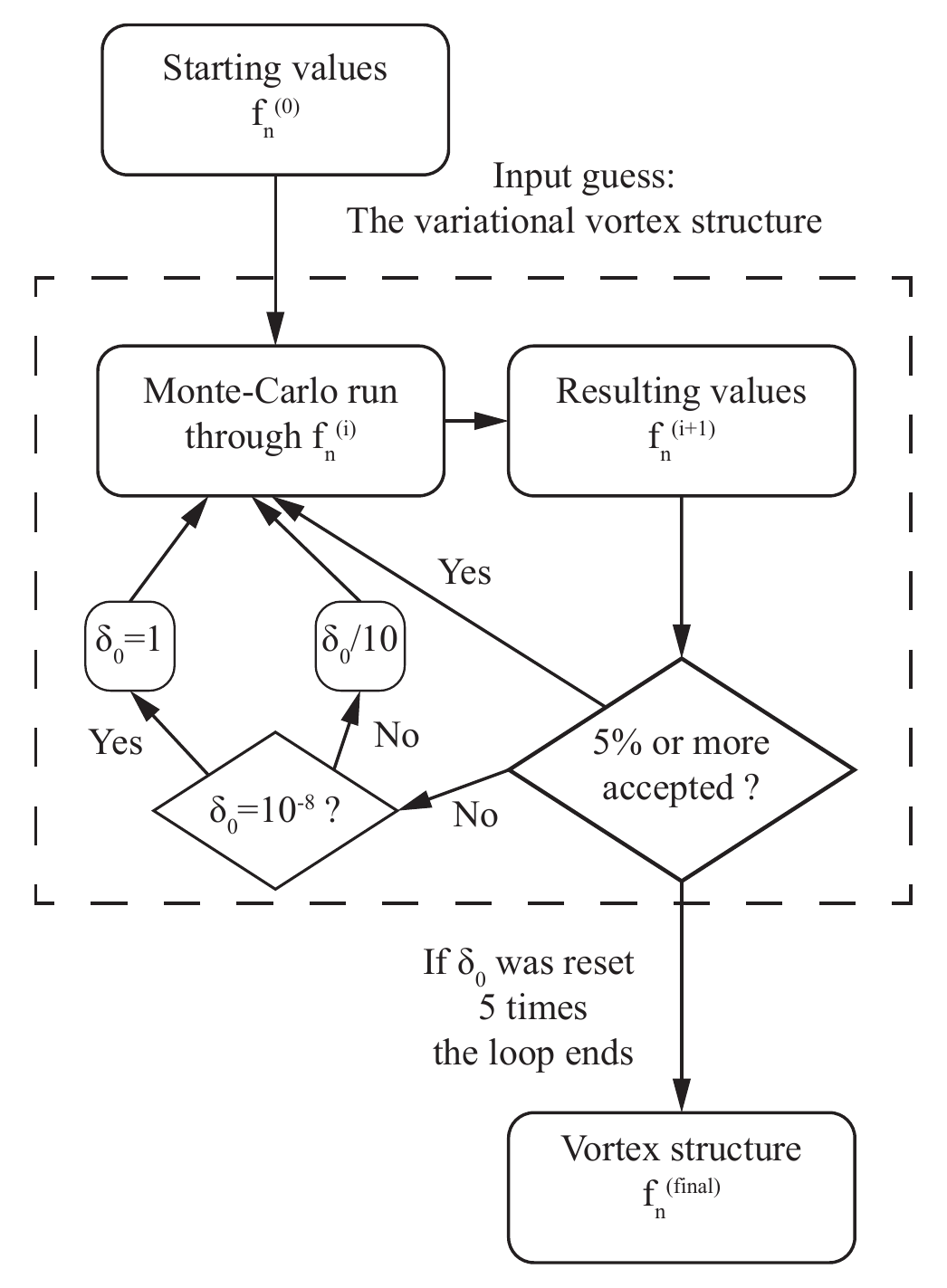}
\caption{The algorithm for the calculation of \textit{one} vortex structure.}
\label{fig:Algorithm}
\end{figure}

The starting value for $\delta_0=1$ (or $m=0$) and this value is lowered (or $m$ is raised) throughout the different loops in the algorithm. The criterion for lowering the value of $\delta_0$ is that 5\% or less of the points $f_n^{i,\pm}$ are accepted, hence if the vortex structure practically doesn't change any more. The Monte-Carlo loop will keep on running until $\delta_0=10^{-8}$ and 5\% or less of the points change. In order to allow for simulated annealing the complete loop will run 5 times, resetting the value of $\delta_0$ to 1 each time. This way it is possible to jump out of a local minimum whenever stuck. The resulting algorithm is depicted in figure \ref{fig:Algorithm}.

In order to get reliable values for our results, we will do 5 independent vortex structure calculations, which will independently be analysed. This allows to give an error bar to the results.

%%%%%%%%%%
%Section 4: Discussion of the results
%%%%%%%%%%
\section{Results and discussion}
\label{sec:ResultsDiscussion}
\begin{figure}[!htb]
\centering
\includegraphics[width=0.65\linewidth]{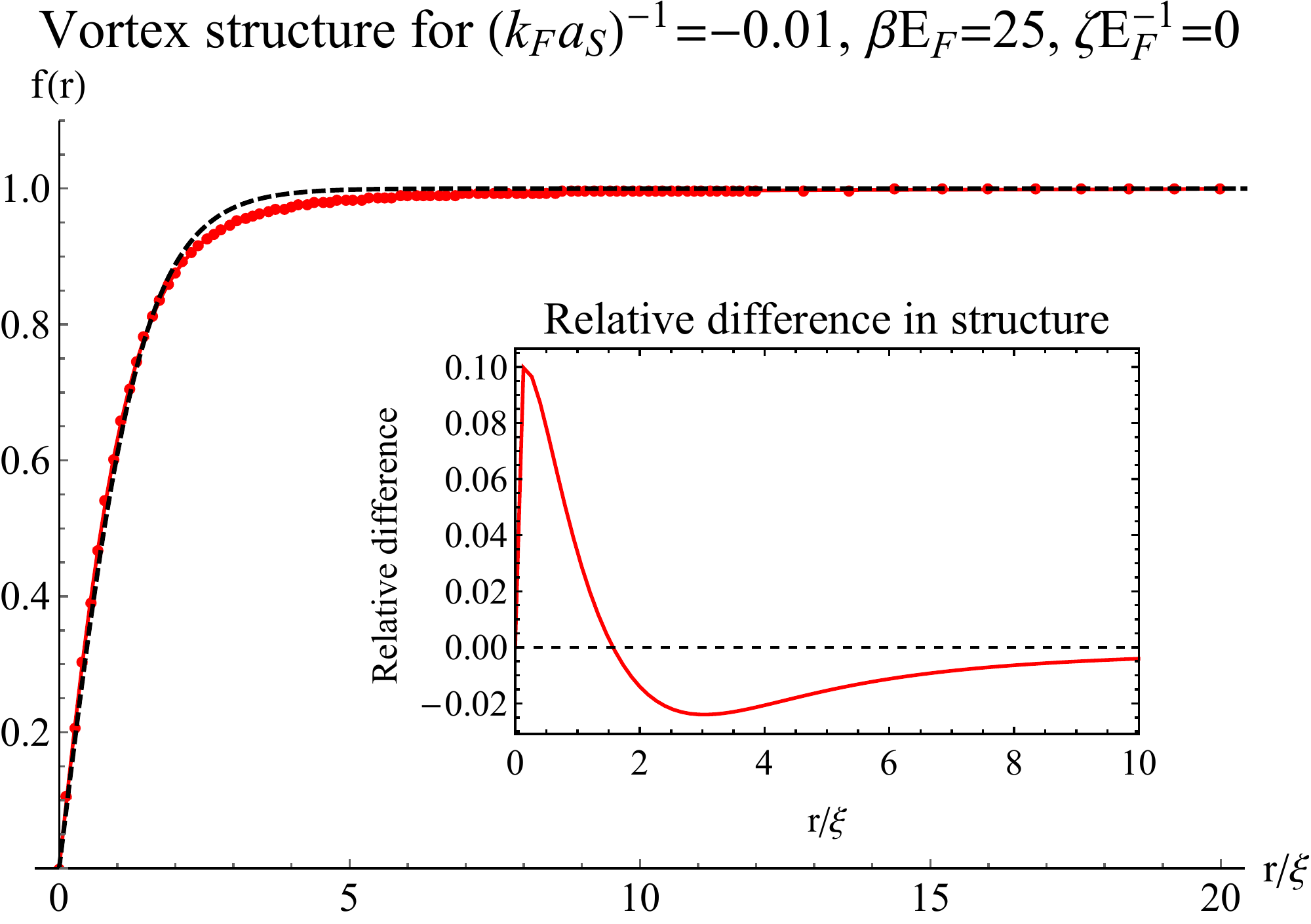}
\caption{The solid red line is the resulting vortex structure for $(a_s,\beta,\zeta)=(-0.01,25,0)$ with error bars. The black dashed line is the variational solution for the same set of parameters. The error bars on the result are so small (about 0.1-0.01\% of the value) that they are not visible on the plot.}
\label{fig:VortexStructure}
\end{figure}

After five runs for each set of values $(a_s,\beta,\zeta)$, a vortex structure is obtained. An example of an obtained structure is given in figure \ref{fig:VortexStructure}, where the red dots show the result. The distribution of the dots also show the discretized grid that was used in the algorithm.
After determining the vortex structure, the following properties of the vortex structure are calculated:
\begin{itemize}
\item The healing length $\xi_\mathrm{num}$, obtained by fitting a tangent hyperbolic to the numerical result.
\item The quadratic distance between the numerical solution $\lbrace f_n^{\mathrm{final}} \left| n\in\mathbb{N}_0\wedge n\leq N+1\right.\rbrace$ and the variational solution with healing length \eqref{eq:HealingLengthSergei}: $\sum_{n=1}^{N+1}||f(r_n)-f_n^{\mathrm{final}}||^2_2$.
\item The goodness of fit, given by $1-R^2$. Being equal to zero in the case of a perfect fit and becoming larger (maximum 1) the worse a fit gets. The value of $R^2$ is defined as the ratio of the model sum of squares to the total sum of squares.
\end{itemize}
For each set of parameters $(a_s,\beta,\zeta)$ the calculation is done five times, leading to a mean value and error (standard deviation). In the subsequent results, only the mean values are shown. The maximum value reached for the relative error was about 1\% for the healing length. The results are discussed in the following subsections.

\begin{figure}[!htb]
    \centering
    
    \begin{subfigure}{\linewidth}
      \centering
      \includegraphics[width=0.8\textwidth]{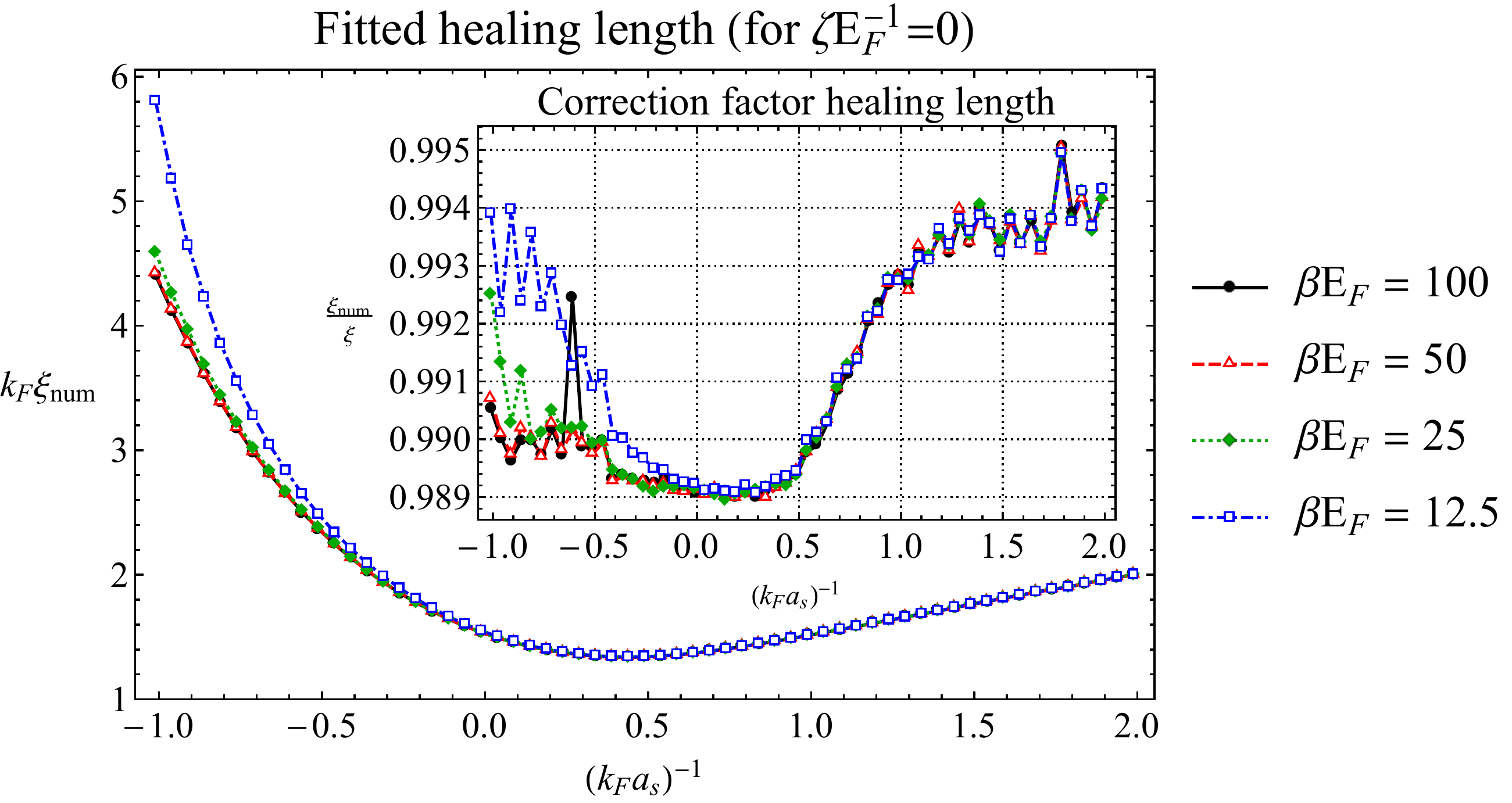}   
    \end{subfigure}
    \begin{subfigure}{\linewidth}
      \centering
      \includegraphics[width=0.8\textwidth]{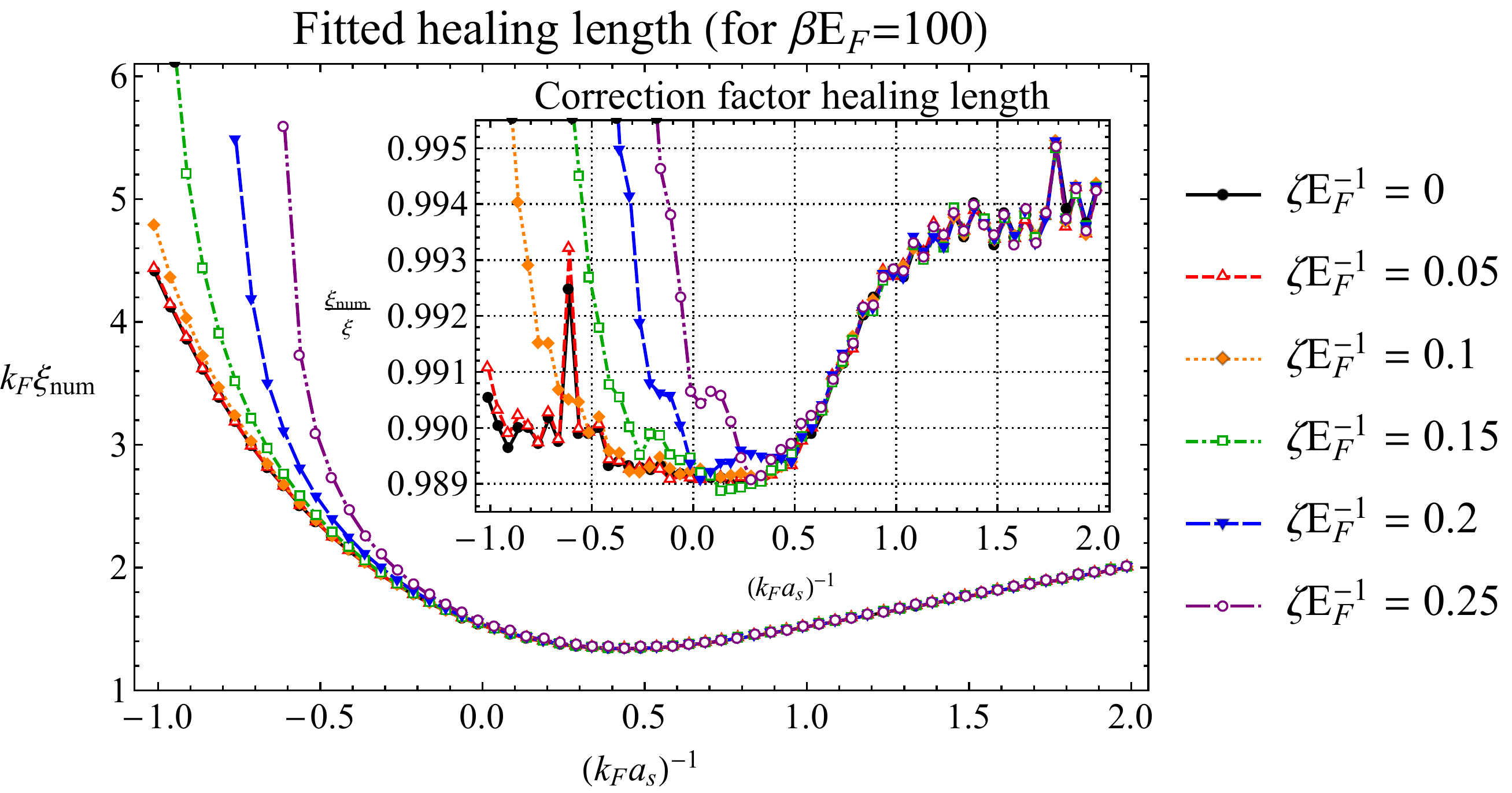}
    \end{subfigure}
     
\caption{The healing length $\xi_{num}$ found by fitting a tangent hyperbolic to the found vortex stucture for different temperatures and polarizations. The inset shows the ratio of the fitted value to the variational value $\xi_{num}/\xi$.}
\label{fig:HealingLengthNumeric}
\end{figure}

%Results for the healing length
In figure \ref{fig:HealingLengthNumeric} the healing length is plotted for different values of the temperature and polarization.

It is obvious from the plots that the correction to the variational healing length \eqref{eq:HealingLengthSergei} is very small. The obtained correction is about 0.1\% to 1.1\% ($\pm 0.2\%$). Only when the polarization becomes large and we go to negative values for $(k_Fa_s)^{-1}$ it is seen that the correction factor suddenly becomes big (up to 10\%). This unstable behaviour is seen in almost all of the results and discussed in the final subsection.

\begin{figure}[!htb]
    \centering
    
    \begin{subfigure}{\linewidth}
      \centering
      \includegraphics[width=0.8\textwidth]{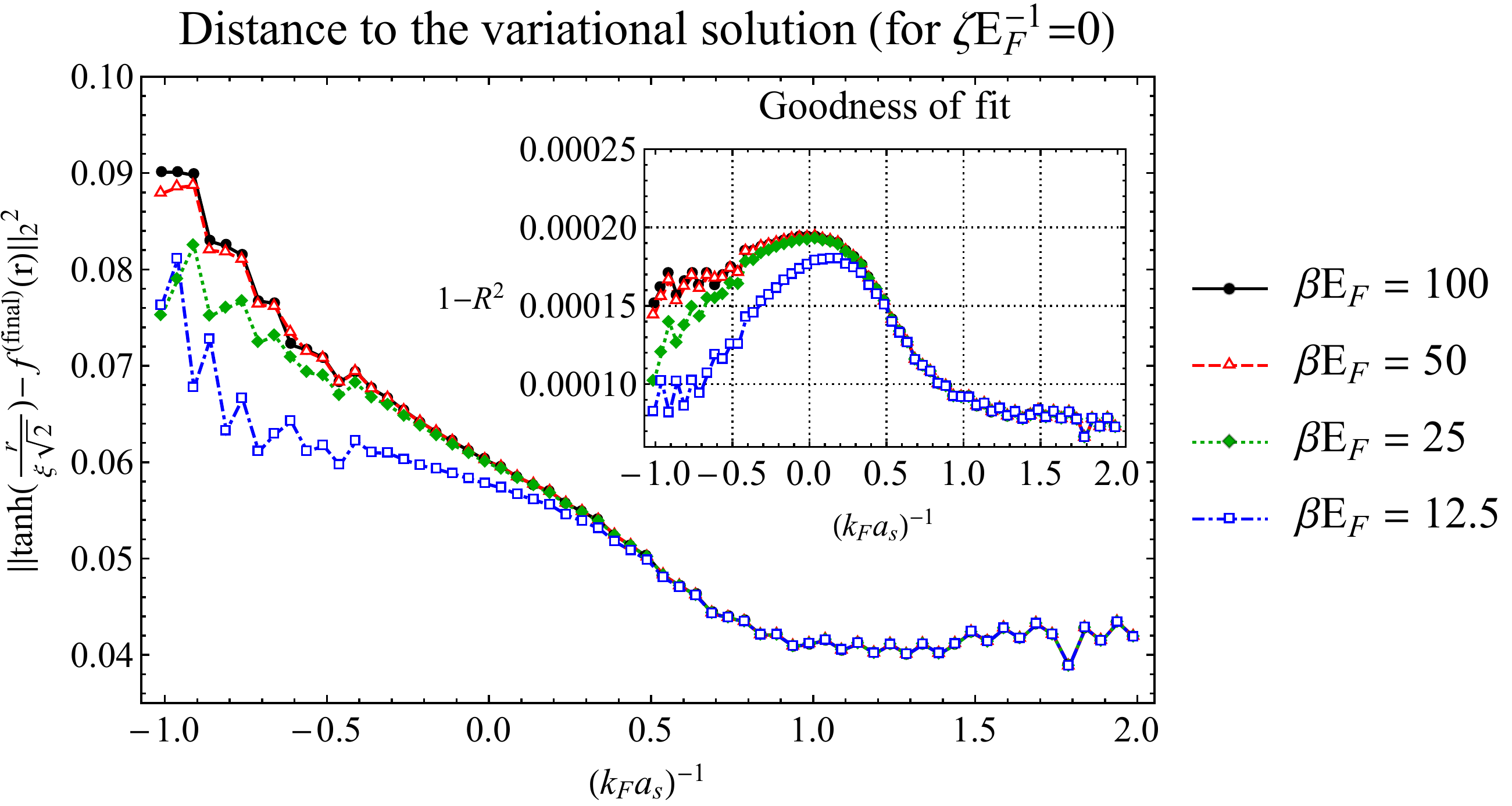}
    \end{subfigure}
    \begin{subfigure}{\linewidth}
      \centering
      \includegraphics[width=0.8\textwidth]{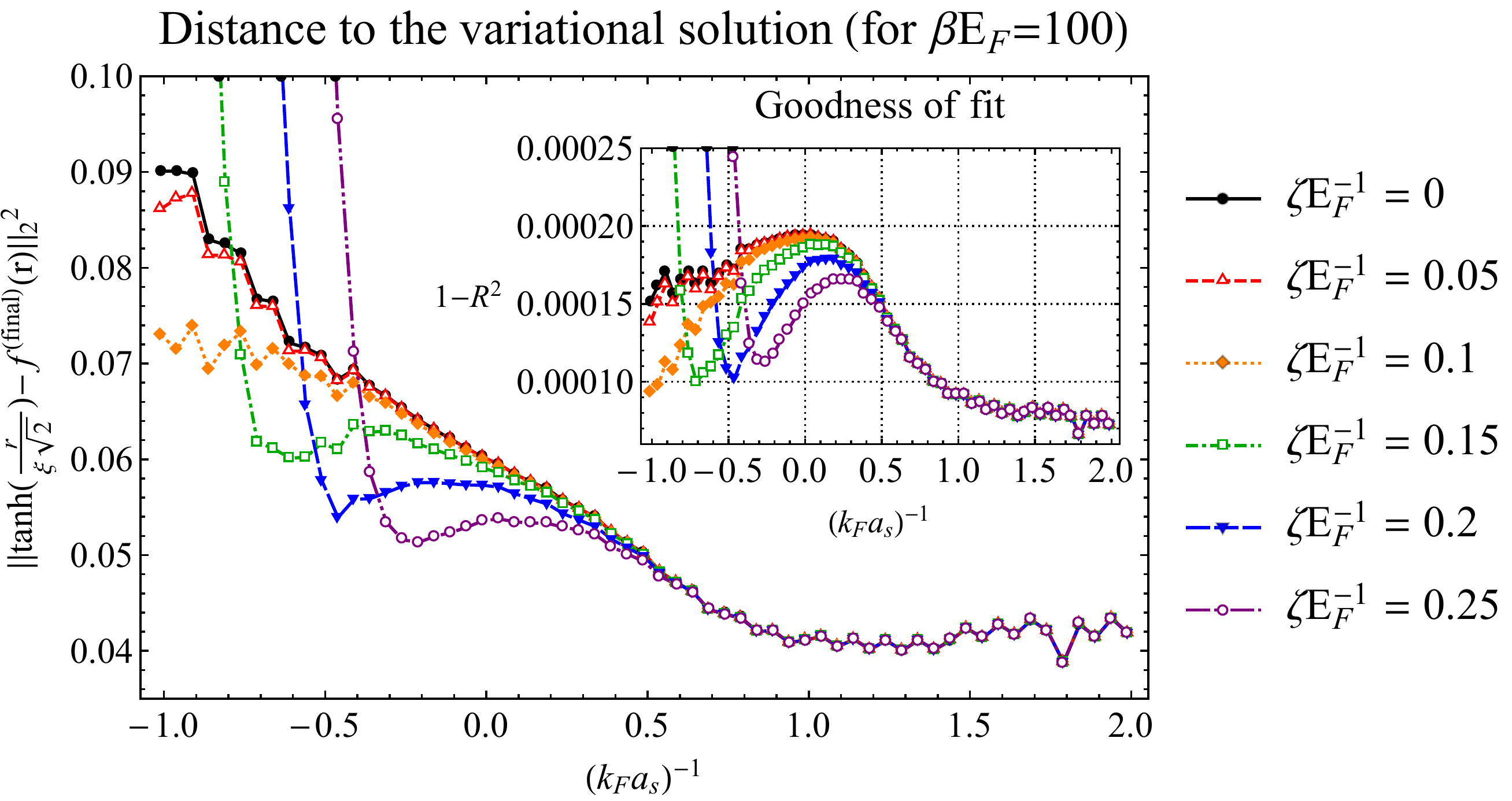}
    \end{subfigure}
     
\caption{The quadratic distance between the variational hyperbolic tangent and the found vortex structure for different temperatures and polarizations. The inset shows the goodness of fit, given by the value $1-R^2$.}
\label{fig:FitNumeric}
\end{figure}

%Results for the goodness of fit
The goodness of fit is determined by looking at the $R^2$-value, together with the square distance between the variational and numerical solution. In figure \ref{fig:FitNumeric} the results are given for different values of the temperature and polarization.

As can be seen the distance between the variational and numerical solution is rather small. Moreover the value of $1-R^2$ is very small, implying that the shape of the variational solution fits the vortex structure rather well.

\begin{figure}[!htb]
    \centering
    
    \begin{subfigure}{\linewidth}
      \centering
      \includegraphics[width=0.8\textwidth]{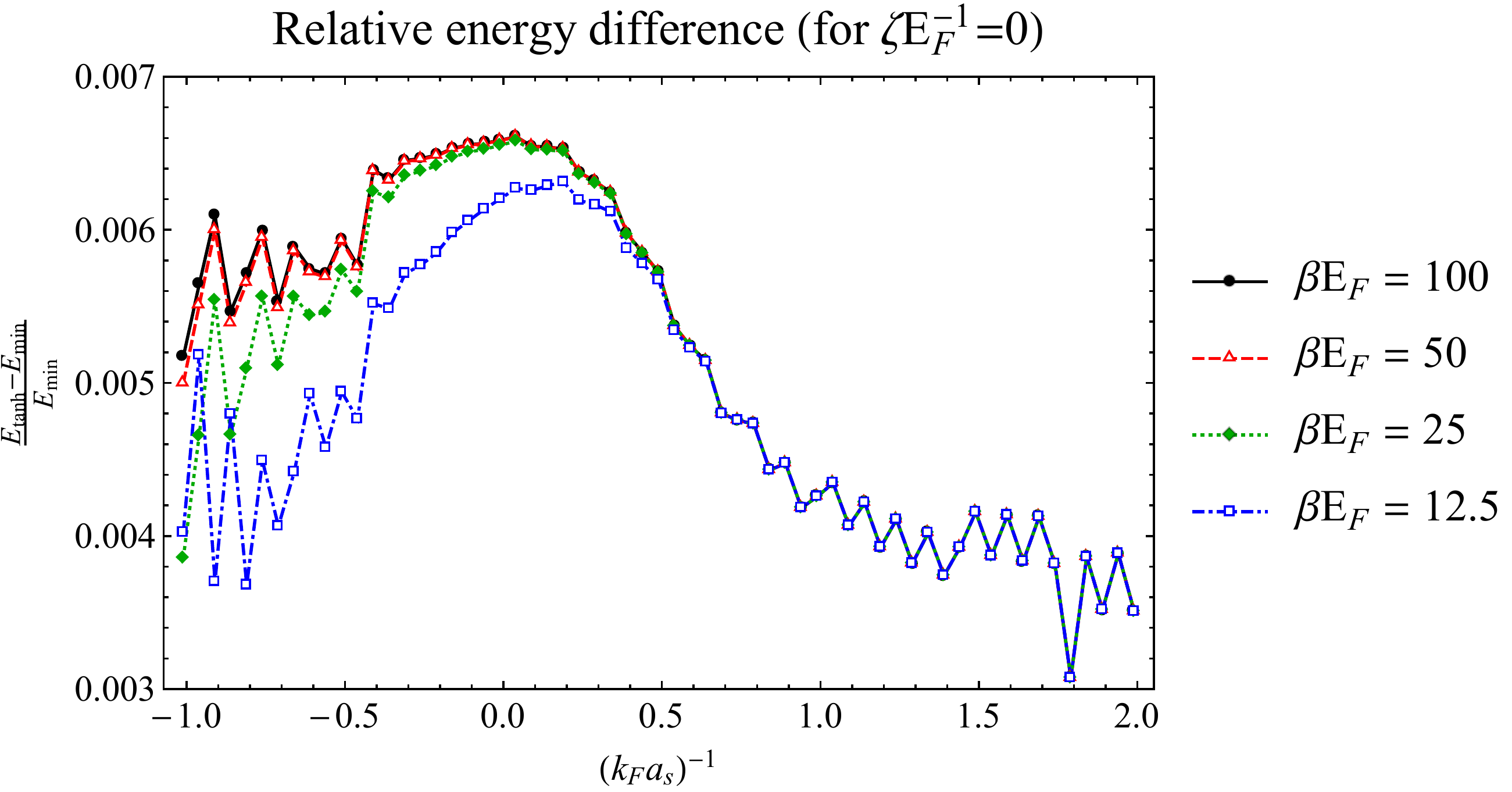}
    \end{subfigure}
    \begin{subfigure}{\linewidth}
      \centering
      \includegraphics[width=0.8\textwidth]{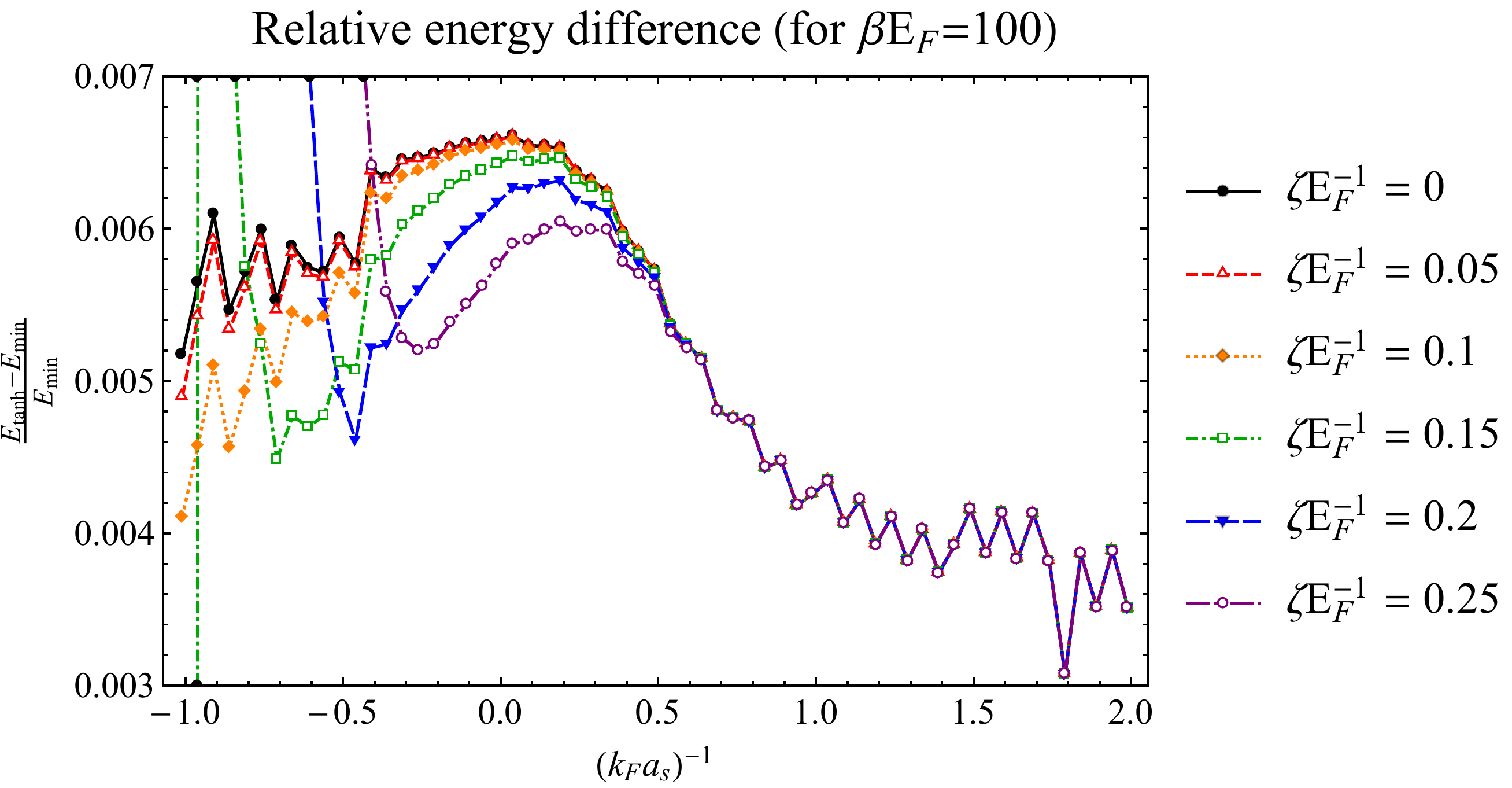}
    \end{subfigure}
     
\caption{The relative energy difference between the variational and found vortex structure for different temperatures and polarizations.}
\label{fig:EnergyDifferenceNumeric}
\end{figure}

%Results for the energy difference
Finally the difference in energy is studied as a function of the scattering length for different temperatures and polarizations. The result can be found in figure \ref{fig:EnergyDifferenceNumeric}.

Comparing the relative energy difference, it can be seen that for most cases (ignoring the unstable behaviour) we see a relative energy difference of about 0.3\% to 0.7\%. This is a clue that the tangent hyperbolic might be sufficient to study (single) vortex behaviour.

\begin{figure}[!htb]
\centering
\includegraphics[width=0.8\linewidth]{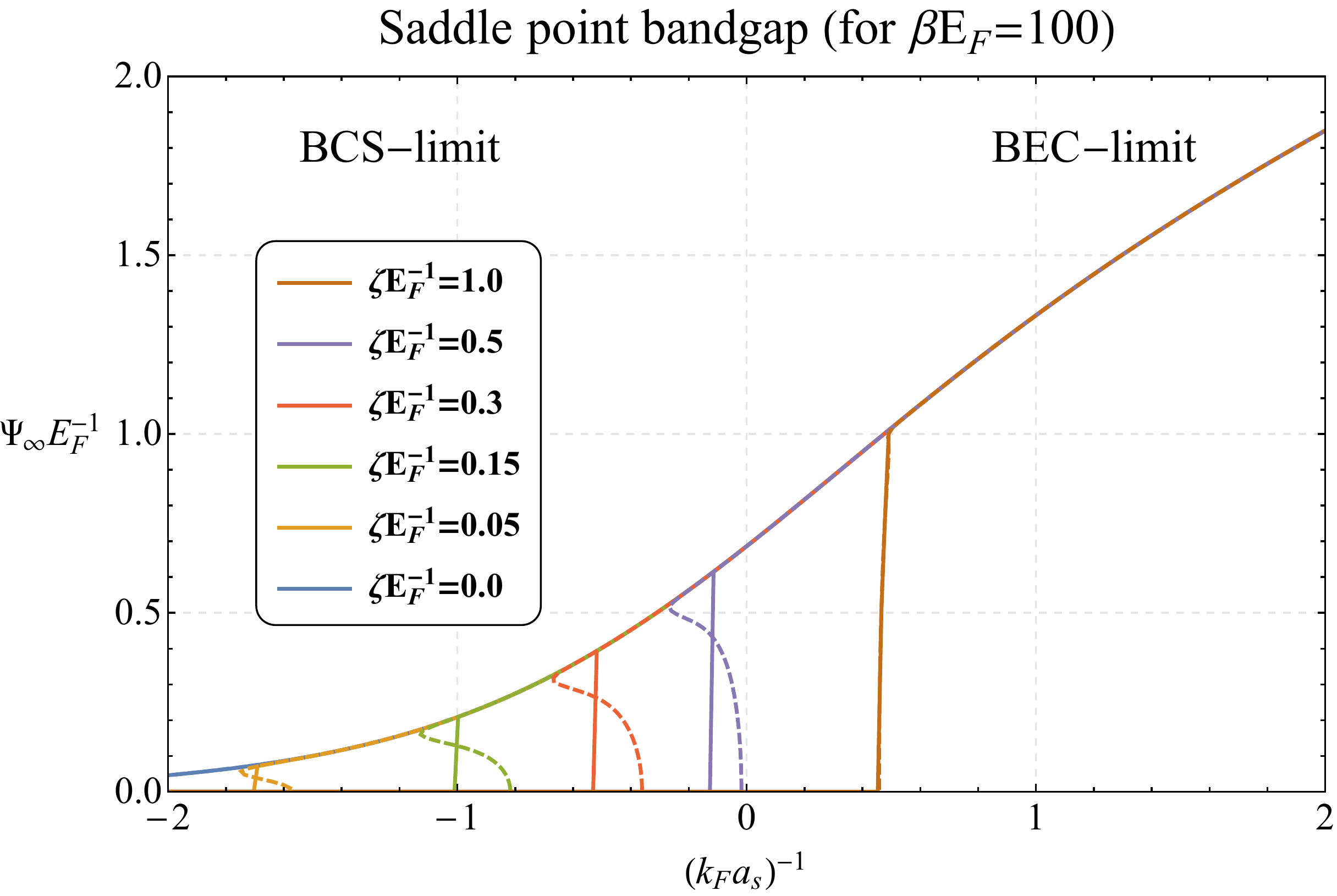}
\caption{The saddle point bandgap $\Psi_\infty$ for $\beta=100$ and different polarizations. The solid line yields the true value for the bandgap (which minimizes the free energy on saddle-point level), the dashed line yields the second (local) minimum.}
\label{fig:SaddlePointBandGap}
\end{figure}

%About the unstable behaviour
For high polarizations at low values of $(k_Fa_s)^{-1}$ it is apparent that there is a turning point where the results became unstable. The reason for this is that in this case the bulk-value $\Psi_\infty$ in equation \eqref{eq:BosonicPairField} approaches zero, so the system is near the so-called Clogston limit, where the imbalance drives a superfluid-to-normal transition. In figure \ref{fig:SaddlePointBandGap} the saddle point gap is sketched.

Note that there are cases where the saddle-point equation has two non-zero solutions for the band gap. It is for the values of $(a_s,\beta,\zeta)$ where there are two possible values for $\Psi_\infty$ that the unstable behaviour appears. This means that the hyperbolic tangent variational model is unstable near the superfluid phase transition.

The values for $(k_Fa_s)^{-1}$ for which the unstable behaviour will start to appear can be easily calculated, this is done by looking for the largest value of $(k_Fa_s)^{-1}$ for which two values of $\Psi_\infty$ are possible. For the cases studied in this article, these values are given in the table below. These can be compared with the low-temperature, no polarization case ($\beta=$100,$\zeta=$0) which becomes unstable for $(k_Fa_s)^{-1}=-2.56$.
\begin{center}
	\begin{tabular}{|c|c||c|c|}
		\hline
		$(\beta,\zeta)$ & $(k_Fa_s)^{-1}$ & $(\beta,\zeta)$ & $(k_Fa_s)^{-1}$\\\hline\hline
		(50,0) & -2.16 & (100,0.1) & -1.09\\\hline
		(25,0) & -1.73 & (100,0.15) & -0.82\\\hline
		(12.5,0) & -1.29 & (100,0.2) & -0.63\\\hline
		(100,0.05) & - 1.58 & (100,0.25) & -0.48\\\hline	
	\end{tabular}
\end{center}
The higher the temperature and polarization become, the sooner the unstable behaviour will start.

\section{Conclusion: Range of applicability}
\label{sec:Conclusion}
In this paper the structure of a vortex was studied using the KTD effective field theory. From the obtained results it can be concluded that, away from the Clogston limit, the hyperbolic tangent:
\begin{itemize}
\item Yields a very accurate guess for the vortex healing length.
\item Gives an excellent fit for the vortex structure.
\item Produces a good estimate for the vortex free energy.
\end{itemize}
This means that, as long as the system is not near the Clogston limit where spin-imbalance destroys superfluidity, the assumption of a hyperbolic tangent for the vortex core profile is valid. Using this analytic fit, the thermodynamic properties can be estimated well, and it is  possible to study also multivortex states.

However the treatment of multivortex states requires some caution. Even though the hyperbolic tangent yields an accurate representation of the vortex structure, there is still some error in the result. This means for example that when one studies single- or multi-vortex states near unitarity, it is impossible to distinguish between two states that only differ slightly (about 1\%) in energy.

\section*{Acknowledgements}
We acknowledge the various fruitful discussions with G. Lombardi, J.P.A. Devreese and W. Van Alpen. This research was supported by the research fund of the University of Antwerp, project: 2014BAPDOCPROEX167 and FFB150168, and by the Flemish Research Foundation (FWO-Vl), project nrs: G.0115.12N, G.0119.12N, G.0122.12N and G.0429.15N.

\end{document}